\documentclass[a4paper,11pt]{article}
\usepackage{jcappub} 
\pdfoutput=1

\usepackage{graphicx}
\usepackage{amsmath,amssymb}
\usepackage{bm}
\usepackage{physics}
\usepackage{booktabs}
\usepackage{multirow}
\usepackage{subcaption}
\usepackage{float}
\usepackage{enumitem}
\usepackage{color}
\usepackage{comment}
\usepackage{diagbox}

\title{Baryonic Feedback across Halo Mass: Impact on the Matter Power Spectrum}

\author[a,1]{Kyle Miller,\note{Corresponding author}}
\author[b]{Surhud More,}
\author[a]{Bhuvnesh Jain}

\affiliation[a]{Department of Physics and Astronomy, University of Pennsylvania, Philadelphia, PA 19104, USA}
\affiliation[b]{Inter-University Centre for Astronomy and Astrophysics, Pune, Maharashtra 411007, India
} 

\emailAdd{mky@sas.upenn.edu, surhud@iucaa.in, bjain@physics.upenn.edu, }

\abstract{

Upcoming weak-lensing surveys will probe the matter distribution at a few percent level on nonlinear scales ($k>1\,{\rm h\,Mpc}^{-1}$) where baryonic feedback from galaxy formation modifies the clustering of matter. Using the IllustrisTNG hydrodynamical simulations, we quantify the mass and radial dependence of baryonic suppression of the matter power spectrum by selectively replacing matter around the center of halos out to a specified radius in the collisionless run with that around their full-physics counterparts. We find that group-scale halos with $\log M_{\rm 200m}/h^{-1}M_\odot \in[13, 14]$ dominate the suppression, contributing a large fraction of the total reduction in power at $k\sim2-30\,h\,{\rm Mpc}^{-1}$, with smaller suppression on either sides of this mass bin. Correctly reproducing the full suppression of the power
spectrum requires accounting for matter redistribution (while enforcing
mass conservation) beyond the virial radius of each halo. We show that the same group-scale regime produces the most detectable deviations in the weak gravitational lensing of background galaxies measured around foreground galaxy groups binned by mass or richness. Such a lensing signal could be a powerful observational test of feedback models together with SZ measurements. Our results motivate emulators that jointly predict the matter power spectrum and halo–matter cross-power spectrum including baryonic effects, enabling unbiased cosmological inference from small scales.
}


\begin{document}
\maketitle
\flushbottom

\section{Introduction}

Weak gravitational lensing is a powerful probe of cosmic structure formation. It enables a direct measurement of the matter distribution in the Universe, without relying on biased tracers such as galaxies or gas. While gravity primarily governs the large-scale evolution of matter, small-scale astrophysical processes can significantly affect this distribution, especially given the statistical precision of ongoing and upcoming weak lensing surveys. In particular, baryonic feedback processes associated with galaxy formation and evolution can modify the matter distribution on small scales.

Mechanisms such as gas ejection by supernovae and active galactic nuclei (AGN) are key drivers of these baryonic effects \cite{Acevedo_2024, Biernacki_2018, Chisari_2019}. Gas cooling and galaxy formation tend to pull baryons deeper into halos, thereby enhancing the small-scale power in the matter power spectrum. Conversely, feedback from star formation or AGN can expel baryons from galactic centers, reducing small-scale power. The ejection of baryons also results in a 'back reaction' where dark matter expands out to larger scales due to the change in gravitational potential \cite{Duffy_2010, van_Daalen_2011, Velliscig_2014}. These opposing mechanisms compete: cooling, star formation and associate feedback typically dominate in low-mass halos, whereas AGN feedback is expected to prevail in more massive systems \cite{gebhardt_2023, Schaye_2014, Biernacki_2018, Nicola_2022}.

Accurate cosmological inference from weak lensing observables therefore requires careful marginalization over baryonic effects. However, this task remains difficult due to the complex and poorly understood nature of baryonic physics \cite{Chisari_2019}. Marginalizing without informative priors from complementary observables can substantially degrade the statistical power of small-scale measurements \cite{pandey_2025,Chen_2022}. Moreover, much of the weak lensing signal-to-noise arises from nonlinear scales where baryonic processes are most influential, complicating interpretation. Implementing conservative scale cuts to avoid such nonlinear regimes limits the constraining power of next-generation surveys, and as survey precision improves, the required scale cuts would need to increase further unless baryonic effects are modeled.

Consequently, developing robust models for baryonic effects on small scales is essential. Numerous studies have attempted to quantify these effects using hydrodynamical simulations (e.g. \cite{van_Daalen_2011, Acevedo_2024, Semboloni2012, Zentner2012, vanloon_2023,vandaalen_2025}). Yet, modeling baryonic feedback remains challenging due to the wide dynamical range of relevant astrophysical processes. Running cosmologically representative simulations at sufficiently high resolution is computationally demanding, and most simulations rely on subgrid prescriptions to model unresolved physics. Differences in these prescriptions can lead to significant variations in predicted outcomes, underscoring the need for continued refinement and cross-validation of baryonic feedback models.

Improved halo models have been developed to mitigate the impact of baryonic feedback on cosmological inference. These models redistribute matter within halos, either in total or by decomposing it into stellar, gaseous, and dark matter components, using density profiles motivated by hydrodynamical simulations \cite{Zentner2008, Zentner2012, Semboloni2012, Semboloni2013, Mead2020, Osato2022,vandaalen_2025}. Such models typically include several free parameters that must be calibrated against simulations or marginalized over in cosmological analyses. While this approach can reduce biases in inferred cosmological parameters, it introduces a statistical penalty and potential model dependence if priors from simulations are imposed. Some of these parameters, particularly those related to gas physics, can in principle be constrained using observational data, for example, through X-ray data or the various correlations of the thermal and kinetic Sunyaev–Zel’dovich (SZ) effects. However, these constraints are often degenerate with cosmological parameters \cite{Osato2022, Nicola_2022, pranjal_2024}.

Recent efforts to constrain baryonic effects directly from observations such as tSZ and kSZ cross-correlations and weak lensing, have primarily focused on feedback signatures in group- and cluster-scale halos ($M_{\rm 200m}>10^{13}M_{\odot}$) (e.g. \cite{To_2024, hadzhiyska2025, Huang_2021, Schneider_2022, bigwood_2024, pandey_2025, xu_2025, siegel_2025}). 
Current cosmic shear measurements extending to small scales have shown conflicting claims about the evidence for baryonic feedback effects. For example, \cite{terasawa2025} show that cosmological inference is unaffected even after including information from small scales and using collisionless predictions to fit the data. This could presumably be due to degeneracies in the baryonic feedback and other systematic effects. However, \cite{Sarmiento_2025} carry out a joint analysis of galaxy and CMB lensing and suggest the presence of some suppression of the power spectrum on small scales. On the other hand, \cite{siegel_2025} claim a large suppression in the matter power spectrum due to baryonic feedback effects in a joint analysis of X-ray, kinetic SZ and weak lensing analysis of groups and galaxy clusters. This motivates the development of new observables and joint inference frameworks capable of breaking degeneracies in a data-driven manner, enabling simultaneous inference of cosmological parameters and baryonic feedback effects.

Fast Radio Bursts (FRBs) have also recently emerged as a promising and complementary observational probe of baryonic feedback. FRBs are extremely energetic milisecond-duration pulses of radio waves of uncertain origins propagating from cosmological distances \cite{Lorimer_2024,Thornton_2013}. The dispersion measures of FRBs provide a direct measurement of the column density of free electrons along a line of sight and are therefore sensitive to the distribution of ionized gas in halos and the circumgalactic and intergalactic media \cite{medlock_2025}. FRB dispersion measures have been used to probe the impact of baryonic feedback on structure formation (e.g. \cite{Baptista_2024, Medlock_2024a}), and have been shown by \cite{medlock_2025} to be tightly correlated with baryon spread, a key metric of the displacement of gas due to baryonic effects. With rapidly advancing observational capabilities, such as CHIME \cite{Leung_2021}, CHORD \cite{Vanderlinde_2019}, DSA-2000 \cite{hallinan_2019}, and BURSST \cite{Lin_2022}, FRB dispersion measures, jointly analyzed with weak lensing and SZ maps, will provide a powerful avenue for constraining baryonic effects and their impact on the matter power spectrum.

In this work, we use the \textsc{IllustrisTNG-300} simulation suite to quantify the contribution of halos in different mass bins to the suppression of the matter power spectrum. By matching halos between the collisionless and full-physics simulations, we construct hybrid snapshots in which selected gravity-only halos are replaced by their baryonic counterparts. We then measure the resulting suppression of the matter power spectrum and examine how this suppression manifests in the group–galaxy lensing signal. This approach allows us to isolate the impact of baryonic feedback across both halo mass and spatial scales. Finally, we discuss how combining cosmic shear and group–galaxy lensing, especially when binned by group properties, can help disentangle baryonic effects from cosmological parameters.

A similar investigation was carried out in \cite{vanloon_2023} with the help of the cosmo-OWLS \cite{Le_Brun_2014} and BAHAMAS \cite{McCarthy_2016} simulation suites. Although our methodology is different in its implementation details, our conclusions are largely consistent with their results, and expand upon them by directly quantifying the impact that the redistribution of matter within a halo due to baryonic feedback effects have on the total suppression of the matter power spectrum, and by exploring baryonic feedback effects across a wider range of spatial scales, including those beyond the virial radius. We compare our findings in section \ref{discussion}.

Throughout this work, we will assume that halo masses are defined as $M_{\rm 200m}$, i.e., within a spherical boundary that encompasses 200 times the mean matter density of the Universe. We will use comoving coordinates to describe the scales involved.

\section{Data} \label{Data}

\subsection{Cosmological Simulations}

The IllustrisTNG simulation \citep{Nelson2019, Pillepich2018,Marinacci2018,Naiman2018,Springel2018} 
is a suite of gravo-magnetohydrodynamic simulations, built with an improved treatment of baryonic physics from the original Illustris simulation suite \citep{Vogelsberger2014a,Genel2014,Sijacki2015,Vogelsberger2014b} 
using the moving-mesh \textsc{AREPO} code \cite{Springel_2010, Weinberger_2020}. A flat $\Lambda$CDM cosmology is used with the parameter values $\Omega_{\rm m}=0.3089$, $\Omega_{\rm b}=0.0486$, $\sigma_8=0.8159$, $n_{\rm s}=0.9667$ and $H_0=67.74\, {\rm km \, s^{-1} Mpc^{-1}}$. 

We focus on the TNG300-1 (full-physics) and TNG300-1-Dark (gravity-only) runs, which share identical initial conditions. These simulations have been carried out in a box size of 205 $h^{-1}$ Mpc, sufficient to capture large-scale modes relevant for our analysis.
The full-physics simulation comprises an initial set of $2\times 2500^3$ particles, half of which are dark matter with a mass resolution of $4.0\times10^7 h^{-1}M_\odot$, and the rest are gas tracer particles with an average cell mass of $7.4\times10^6 h^{-1}M_\odot$, the minimum gas softening length of $370$ pc, and a physical softening length for dark matter equal to $1.48$ kpc. The gravity-only simulation does not have the gas tracer particles, assuming baryons to be collisionless and to evolve only under the effect of gravity. The public data release provides 99 snapshots between $z=20$ and $z=0$. 

The halo catalog is constructed using the friends-of-friends group-finding algorithm with subhalos identified with the \textsc{Subfind} algorithm \citep{Springel2001,Dolag2009}.
We use the bidirectional matching of LHaloTree identified subhalos \citep{Nelson_2015} to cross-identify halos between TNG300-1 and TNG300-1-Dark. Subhalo merger trees are constructed using the LHaloTree algorithm by linking subhalos to their descendants across simulation snapshots based on the number of shared particles of all types, weighted by particle binding energy. By emphasizing the most tightly bound (inner) particles, this method remains robust to tidal stripping, baryonic feedback, and infall into larger host halos, all of which can strongly impact subhalo boundaries and redistribute particles in the outskirts of halos. Subhalos in the full-physics and the gravity-only simulation are matched bidirectionally by identifying pairs with the largest number of shared dark matter particles. Since only dark matter particles are used, subhalo matching is not directly impacted by phase transitions (e.g. gas to stars).

We consider halos to be a match if they share the same \textsc{firstSub} subhalo from the \textsc{Subfind} algorithm, defined as the most massive subhalo within the FoF group. This matching scheme is chosen because \textsc{firstSub} subhalos are generally central halos within FoF groups, making it rare ($<1\%$) for matched \textsc{firstSub} subhalos to belong to different parent halos or for a matched pair to correspond to a central subhalo in one simulation and a satellite in the other, particularly for high-mass halos. 

Approximately $\sim 3\%$ of halos in our sample lack a bijective LHaloTree match between the full-physics and dark-matter-only simulations and are therefore excluded from our analysis. The vast majority ($97\%$) of these unmatched systems lie in the lowest mass bin ($\log_{10}(M_{\rm 200m}/h^{-1}M_\odot) \in [12, 13]$). These halos typically either lack a well-defined \textsc{firstSub} subhalo or contain too few particles in the \textsc{firstSub} subhalo to permit a reliable bidirectional match. This unmatched set includes systems that are strongly affected by feedback, mergers, or tidal disruptions, potentially biasing our results with their exclusion if significant baryonic feedback processes are ignored in lower mass bins. However, we expect any such bias to be minimal for several reasons: (i) unmatched halos constitute only a small fraction of the total population, (ii) matching failures can arise from multiple causes beyond baryonic feedback alone, and (iii) the unmatched systems are overwhelmingly low-mass halos, which contribute less to the total mass budget and aggregate mass redistribution. Additionally, repeating our analysis using the unidirectional SubLink algorithm for matching, which yields roughly double the number of unmatched halos, produces nearly identical results, indicating that any potential bias from excluding these unmatched halos is negligible.

The friends-of-friends halo finder is able to account for the asphericity in the distribution of matter, but can select halos which differ in their spherical overdensities in a concentration and mass dependent manner \citep{More2010}. Given that baryonic effects can also change the halo triaxiality including the orientation, we decide to use spherical overdensity mass estimates in our analysis.

In our primary analysis, we will restrict ourselves to the $z=0$ snapshot for simplicity, where baryonic feedback effects are most pronounced. However, we also replicate a subset of our analysis with the $z=0.5$ snapshot (see Appendix~\ref{redshift}), which is most relevant for weak lensing analyses. We will only analyze halos with mass $M_{\rm 200m} \geq 10^{12}\,h^{-1}M_\odot$, encompassing galaxy- to cluster-scale halos, to focus on halos that are well resolved in simulations. We primarily focus on mass bins up to $10^{14.5}\,h^{-1}M_\odot$, as bins above this threshold contain very few halos. 

\section{Methodology} \label{Methodology}

Our goal is to quantify the impact of baryonic effects on the matter distribution, specifically on the matter and galaxy–matter correlation functions. To directly assess this impact as a function of halo mass and radial extent, we construct a series of hybrid matter distributions by selectively replacing particles in halos of the TNG300-1-Dark simulation with their counterparts from the full-physics TNG300-1 run.

\subsection{Isolating baryonic effects by halo mass}

We consider halos in logarithmically spaced mass bins spanning $\log_{10}(M_{\rm 200m}/h^{-1}M_\odot) \in [12, 14.5]$. The use of spherical overdensity (SO) halos enables particle replacement within equal-volume regions in the gravity-only simulation, avoiding complications arising from potential shape mismatches between corresponding halos in the two runs. Because the centers of matched halos do not coincide exactly between the simulations, replacements are performed relative to the assigned halo center in each matched pair.

For each mass bin, we identify matched halos between the full-physics and gravity-only simulations using the subhalo matching method outlined in section \ref{Data}, ignoring for replacement any halo with no clear match. Around the center of each matched halo, we define a spherical replacement region of radius $R = \alpha R_{200\mathrm{m}}$, where $\alpha \in {0.5, 1.0, 1.5, 2.0}$. Within this region, we replace all particles in the gravity-only simulation with those from the full-physics simulation—including dark matter, gas, stars, and black holes—while preserving both particle positions relative to the halo center and individual particle masses. To avoid double counting, we do not replace particles in smaller halos whose replacement regions overlap with those of more massive halos that are also being replaced. This is equivalent to adopting a hard halo exclusion scheme, where halos are not allowed to overlap. 

For small replacement radii, $\alpha \in [0.5,1.0]$, at most 5\% of halos are excluded in any combination of mass bins. With larger replacement radii, $\alpha \in [1.5,2.0]$, exclusion is more significant, on the order of 10\% for most mass bin combinations, and $\sim20\%$ for the case when all bins are replaced. While it may seem that many particles are being ignored in this scheme, most of these exclusions occur because one halo hosts another, especially when several mass bins are replaced at once. Exclusion does not affect the replacement of these particles that are shared by the halos, so only in edge cases of partial overlap there could be some particles which have not been replaced.

Hard exclusion likely leads to a small underestimate in the matter power suppression for large replacement radii. However this is necessary to avoid more complications. For example, halo centers are often offset between the full-physics and gravity-only simulations, requiring that particle locations be offset by the same amount, however, if a particle is shared by multiple halos, choosing its parent arbitrarily decides its offset and will result in artificial gaps or clumps in the distribution of particles at the boundary between halos.

A further complication arises because matched halos in the full-physics and gravity-only simulations may not have identical $M_{\rm 200m}$ values or enclosed masses within the replacement radius, owing to baryon-driven mass redistribution. Simply replacing regions with lower total mass can introduce a spurious increase in the large-scale correlation amplitude. This occurs because a reduction in enclosed mass leads to fewer total particles in the box and hence fewer random pairs at larger separations, artificially boosting the correlation on large scales. This effect, typically at the few-percent level, is most pronounced for smaller replacement radii ($\alpha \in [0.5, 1.0]$) and diminishes for larger $\alpha$. 
We enforce mass conservation on a halo-by-halo basis, uniformly redistributing the difference in mass incurred by replacement to the surrounding gravity-only simulation particles in a spherical region out to two times the replacement radius. This physically motivated approach conserves each halo mass (and global mass) and mimics the physical reality of baryonic effects ejecting mass to larger scales. 
We explore this method and alternatives in Appendix \ref{masscons}, and demonstrate that enforcing mass conservation successfully removes the spurious large-scale signal. Important to note is that the relative contribution of each halo mass bin to suppression is entirely unchanged by this mass conservation method.

The procedure described above produces a modified simulation snapshot in which only halos within a specified mass bin contain matter redistributed by baryonic feedback, while the remainder of the volume evolves according to gravity alone. This construction enables us to isolate the contribution of baryonic effects as a function of both halo mass and radial distance from the halo center.

By solely quantifying high-mass halo contributions to the suppression of the matter power spectrum, we neglect the effect of baryons in cosmic structures outside of halos, including filaments, walls, and voids, which are expected to contribute subdominantly, but potentially non-negligibly to matter power suppression. \cite{Sunseri_2023} explores the contribution of non-halo structure to suppression using IllustrisTNG. They find that baryonic feedback suppresses matter clustering in halos by $20\%$, compared to only $5\%$ in heavy filaments ($M_{\rm{min}}>10^{13}M_\odot/h$) and significantly less for lighter filaments, walls, and voids. They show that although halos and filaments with mass $M>10^{11}M_\odot/h$ correspond to similar total mass fractions ($\sim40\%$), halos account for nearly all of the suppression of the matter power spectrum, even when only including halos with mass $M>10^{12}M_\odot/h$. We show in section \ref{results} that our results are consistent with these findings, and that the contribution by filaments is subdominant to the halo contributions.

\subsection{Measurements of the matter power spectrum}

To quantify the suppression induced by baryonic feedback for a given halo mass bin, we compute the matter power spectrum of each modified snapshot using the Fast Fourier Transform (FFT) algorithm implemented in \textsc{Nbodykit} \citep{Hand2018}, weighting by particle mass. We evaluate all unique combinations of halo mass bins and $\alpha$ values, as well as all pairwise combinations of two different mass bins and $\alpha$, to identify how interactions between halos of different masses contribute to changes in the matter power spectrum. 

We compute the power spectrum up to a wavenumber of $k = 31.32 h\mathrm{Mpc}^{-1}$, using an FFT mesh of size $N_{\rm{mesh}}=512^3$ and utilizing the 'self-folding' technique outlined in \cite{Jenkins1998}. The folding method allows for efficient small-scale power spectrum measurements by mapping the box on onto a factor $f_{\rm{fold}}$ subdivision of itself, essentially accomplishing the same thing as a finer mesh, but at significantly smaller computational memory footprint. We use two foldings of $f_{\rm{fold}}=2$ and $f_{\rm{fold}}=4$, where $f_{\rm{fold}}=2$ covers the power spectrum from $k\sim8h\mathrm{Mpc}^{-1}$ to $k\sim16h\mathrm{Mpc}^{-1}$, and $f_{\rm{fold}}=4$ covers $k\sim16h\mathrm{Mpc}^{-1}$ to $k\sim32h\mathrm{Mpc}^{-1}$. We use two foldings because each extra fold adds more noise to the measurement, so we only include regions where folding extends beyond the $k$ limit of the previous folding to minimize the presence of this noise. Higher precision can be obtained with a finer mesh, but it is not necessary for our analysis.

While IllustrisTNG enables the power spectrum to be computed to much smaller scales ($k\sim125\,{\rm h}\,{\rm Mpc}^{-1}$) \cite{Springel2018}, we include only the range most relevant for weak lensing analysis. At this scale, we are able to measure the point of maximum suppression before baryonic feedback effects are dominated by processes within halos that boost matter power.

We note that our methodology is tailored to identify the impact of the redistribution of matter in halos of a particular mass due to baryonic effects on the total suppression. In a complementary approach \cite{vanloon_2023} directly measure the power spectrum of particles within halos of a chosen mass bin. In our case, we compute the total power spectrum of particles affected by baryonic feedback effects in halos of a given mass together with the rest of the mass which is assumed to follow only gravity, thus isolating the impact of baryonic effects from halos in a single mass bin.

\section{Results} \label{results}

\subsection{Halo mass and radial dependence of the baryonic feedback}

\begin{figure}
    \centering
    \includegraphics[width=\linewidth]{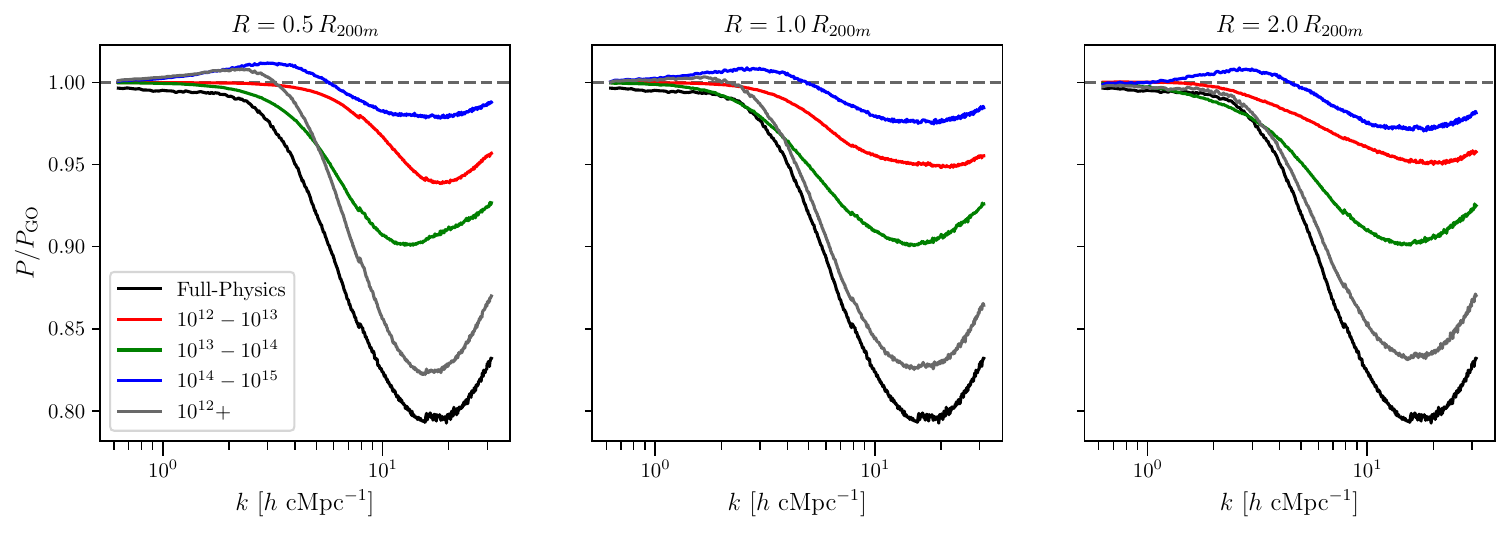}
    \caption{The ratio of the full-physics and gravity-only power spectrum $P/P_{\rm{GO}}$ for each non-overlapping one-dex mass bin. The three panels show three different replacement radii, i.e. the radii out to which matter is redistributed due to baryonic feedback. The gray curve is the power spectrum ratio for the case where all halos above $10^{12}h^{-1}M_{\odot}$ are replaced at once. The black curve is the ratio for the power spectrum computed on the entire full-physics simulation.}
    \label{fig: plots}
\end{figure}

In this section, we quantify the suppression of the matter power spectrum for each replacement configuration by evaluating the ratio $P / P_{\rm GO}$, where $P$ denotes the power spectrum of the modified hybrid matter distribution and $P_{\rm GO}$ corresponds to that from the gravity-only simulation. The three panels of Fig.~\ref{fig: plots} illustrate results for different replacement radii: the left panel shows $\alpha = 0.5$, the middle shows $\alpha = 1.0$, and the right panel shows $\alpha = 2$. In each panel, the solid black line represents the total suppression of the matter power spectrum, calculated as the ratio of the full-physics to gravity-only power spectra.

On large scales ($k < 2\,h{\rm Mpc}^{-1}$), the two power spectra agree to better than one percent, indicating that baryonic effects have minimal impact in this regime. At intermediate scales ($k \in [3, 20]\,h\,{\rm Mpc}^{-1}$), we observe a steady suppression of power reaching up to $\sim20\%$, consistent with the expected influence of baryonic feedback processes. At smaller scales, the suppression plateaus and is followed by a mild upturn in power, reflecting the increasing dominance of baryonic contraction and star formation within halo cores.

\begin{table}[] 
    \centering
    \resizebox{\textwidth}{!}{
    \begin{tabular}{|c|c|c|c|c|}
        \hline
        \diagbox[]{$\log_{10}(M_{\rm{200m}})$}{Radius} & $0.5R_{\rm{200m}}$ & $1R_{\rm{200m}}$ & $1.5R_{\rm{200m}}$ & $2R_{\rm{200m}}$ \\ \hline
        
        $12-12.5$ & 
        2.2 & 1.7 & 1.7 & 1.6 \\ \hline
        $12.5-13$ &
        4.3 & 3.6 & 3.5 & 3.4  \\ \hline
        $13-13.5$ &
        4.9 & 4.8 & 4.8 & 4.8  \\ \hline
        $13.5-14$ &
        5.1 & 5.3 & 5.4 & 5.5  \\ \hline
        $14-14.5$ &
        2.3 & 2.6 & 2.7 & 2.9  \\ \hline \hline
        $\textbf{12-13}$ &
        \textbf{6.2} & \textbf{5.2} & 5.1 & \textbf{5.0}  \\ \hline
        $12.5-13.5$ &
        9.0 & 8.1 & 8.1 & 7.9  \\ \hline
       $\textbf{13-14}$ &
        \textbf{9.9} & \textbf{10.0} & 10.0 & \textbf{9.9}  \\ \hline
        $13.5-14.5$ &
        7.3 & 7.8 & 7.8 & 8.0  \\ \hline 
        $\textbf{14-15}$ &
        \textbf{2.2} & \textbf{2.5} & 2.7 & \textbf{3.0}  \\ \hline \hline
        $12-12.5$ $\&$ $13-13.5$ &
        6.4 & 6.4 & 6.3 & 6.2  \\ \hline
        $12-12.5$ $\&$ $13.5-14$ &
        6.8 & 6.9 & 6.9 & 6.9  \\ \hline
        $12-12.5$ $\&$ $14-14.5$ &
        4.4 & 4.2 & 4.3 & 4.4  \\ \hline
        $12.5-13$ $\&$ $13.5-14$ &
        9.3 & 8.6 & 8.7 & 8.7 \\ 
        \hline
        $12.5-13$ $\&$ $14-14.5$ &
        6.6 & 6.0 & 6.1 & 6.2 \\
        \hline
        $13-13.5$ $\&$ $14-14.5$ &
        7.0 & 7.3 & 7.3 & 7.5 \\ \hline \hline
        $\textbf{12+}$ & \textbf{17.8} & \textbf{17.5} & 17.3 & \textbf{16.9}  \\
        \hline 
    \end{tabular}}
    \caption{The maximum suppression percentage for each combination of mass bin and replacement radius. The table has been separated into half-dex bins, one-dex bins, and non-adjacent half-dex bins. In \textbf{bold} are the combinations plotted in figure \ref{fig: plots}. Also included is the case where all halos above $10^{12}h^{-1}M_{\odot}$ are replaced. The maximum suppression percentage of the entire full-physics simulation is 20.7\%}
    \label{tab:table 1}
\end{table}

\begin{figure}
    \centering
    \includegraphics[width=0.65\linewidth]{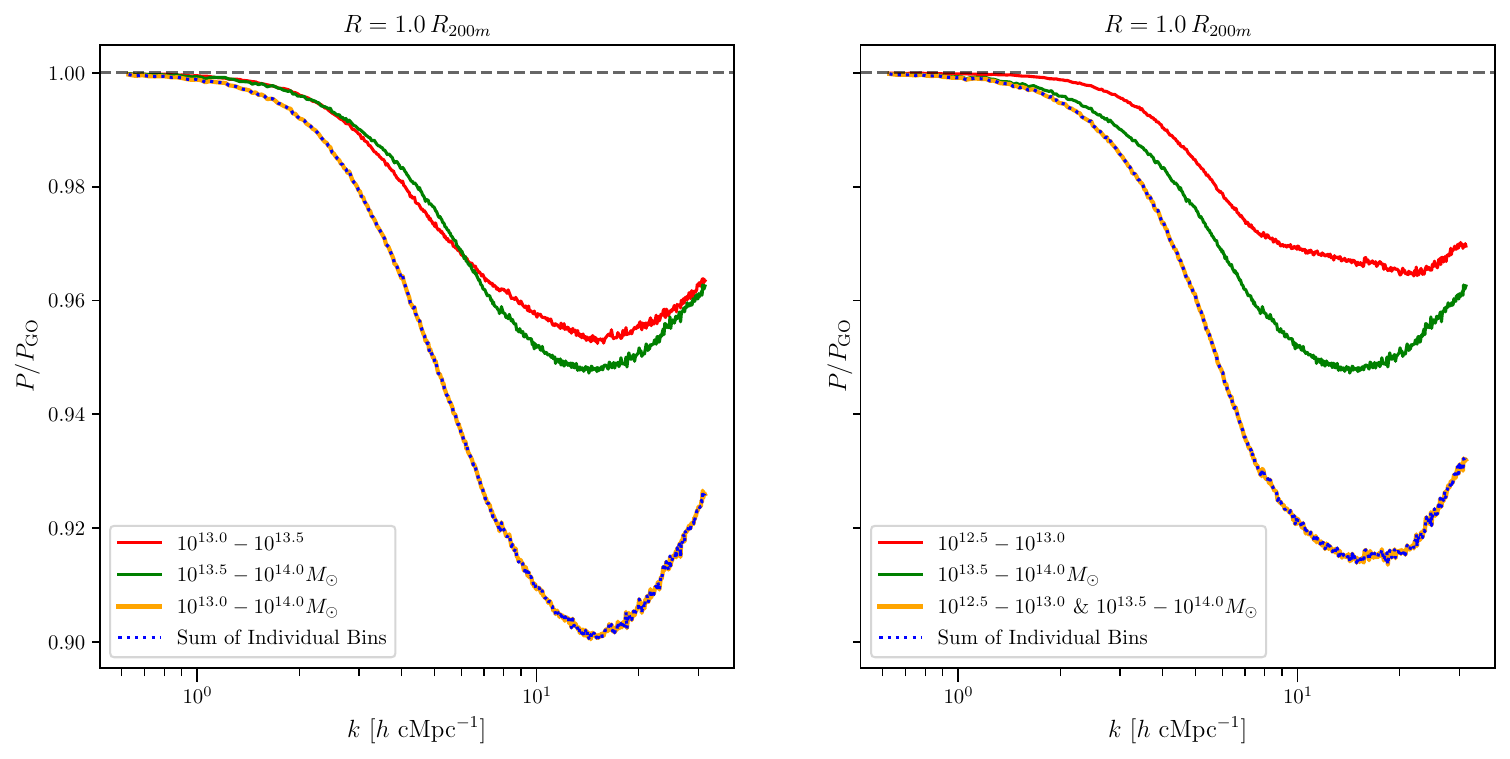}
    \caption{\textbf{Left:} The summation of the power spectrum ratio $P/P_{\rm{GO}}$ for two adjacent half-dex halo mass bins compared to the power spectrum computed from joint replacement of the two bins. \textbf{Right:} The same, but for two non-adjacent half-dex bins. The dotted blue curve is the sum of the red and green curves, and the solid orange curve is from the power spectrum computed with both half-dex bins replaced at once. The two curves are nearly identical in both cases, indicating that there are no significant cross-correlations between mass bins}
    \label{fig:adding bins}
\end{figure}

When particles from halos with $M_{\rm 200m} > 10^{12}\,h^{-1}M_\odot$ in the gravity-only simulation are replaced with those from the full-physics simulation within a region extending to $\alpha R_{\rm 200m}$, we obtain the grey curve in Fig.~\ref{fig: plots}. The redistribution of matter within these halos produces a pronounced suppression in the matter power spectrum. For all $\alpha$, the modification accounts for more than 90\% of the total suppression observed in the full-physics simulation. We expect that this slightly underestimates the true contribution from halos due to the exclusion scheme outlined in section \ref{Methodology}, with the remaining contribution arising from lower-mass halos and from diffuse gas residing beyond the nominal halo boundaries, including in structure such as filaments, walls, and voids. These results are consistent with the findings of \cite{Sunseri_2023}, which finds the contribution of filaments to be subdominant to halos, with walls and voids contributing negligibly.

To further disentangle the relative contributions by halo mass, we divide the sample into three logarithmic mass bins of $\log_{10}(M_{\rm 200m}/h^{-1}M_\odot)$, as shown in Fig.~\ref{fig: plots}. The red, green, and blue solid curves correspond to halos in the ranges $[12,13]$, $[13,14]$, and $[14,15]$, respectively. Low-mass halos ($[12,13]$) are numerous and possess relatively shallow potential wells. The redistribution of matter within these systems produces a moderate suppression of around 5\% in the total matter power spectrum. At the high-mass end ($[14,15]$), halos can host powerful AGN capable of driving large-scale outflows, yet their rarity and their deeper potential wells limits their overall contribution, leading to a modest suppression of only about a couple of percent.

The dominant effect arises from group-scale halos with $\log_{10}(M_{\rm 200m}/h^{-1}M_\odot) \in [13,14]$, which produce a significantly larger suppression of $\sim10\%$. This mass range accounts for $\sim60\%$ of the total suppression attributed to baryonic redistribution in halos above $10^{12}\,h^{-1}M_\odot$. These results highlight the crucial role of feedback processes operating in group-scale halos and its large influence on the matter distribution relevant for interpretation of the cosmic shear.

The halo mass dependence of the impact of the baryonic effects seems to be consistent with the predominance of the fraction of total matter present in different halo mass bins. This can be computed as
\begin{align}
    f(M_1, M_2) = \frac{1}{\bar\rho}\int_{M_1}^{M_2} dM n(M) M\,
\end{align}
where $n(M)dM$ is the halo mass function and represents the number density of halos between $[M, M+dM]$, and $\bar\rho$ is the average matter density of the Universe. These mass fractions are 7.8  percent, 8.4 percent and 4.3 percent in the mass bins [12, 13], [13, 14] and [14, 15], respectively. The effect is thus largest in the mass bin which contributes most to the mass fraction.

\begin{figure}
    \centering
    \includegraphics[width=\linewidth]{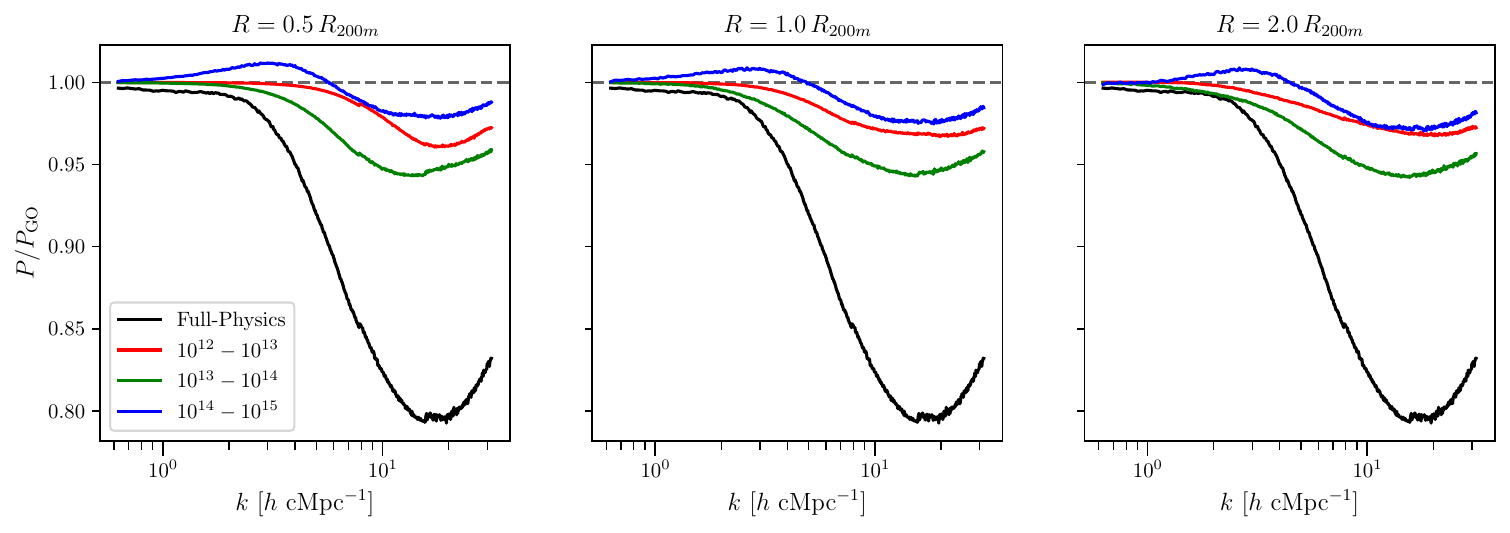}
    \caption{The ratio of the full-physics and gravity-only power spectrum $P/P_{\rm{GO}}$ for each non-overlapping one-dex mass bin, but with halos replaced such that the total mass of replaced halos is the same for all three mass bins. The three panels show three different replacement radii. The black curve is the ratio for the power spectrum computed on the entire full-physics simulation.}
    \label{fig: massfrac plots}
\end{figure}

However, mass fraction alone is insufficient to fully account for group-scale dominance of the suppression. This is clear as group- and galaxy-scale halos make up a similar total mass fraction, but group-scale halos contribute nearly double to overall suppression. To test this explicitly, we repeat our analysis for the mass bins presented in Fig.~\ref{fig: plots} while only replacing halos such that the total mass of replaced halos is the same for each bin. The results of this test are shown in Fig.~\ref{fig: massfrac plots}. By replacing equal total mass fractions we are able to isolate the unique impact of astrophysical processes driving feedback in each mass bin on matter power suppression. We find that group-scales still contribute disproportionately to suppression, nearly doubling galaxy- and cluster-scale halos, which contribute similar amounts. It is clear that the relative difference in suppression contribution comes from both mass fraction and the unique astrophysical properties of each halo population. While group- and galaxy-scale halos make up similar mass fractions, group-scale halos contribute nearly double on a ``per-mass" basis. On the other hand, galaxy- and cluster-scale halos contribute similarly on a per-mass basis, but the cluster-scale halos make up just over half of the total mass fraction of the galaxy-scale halos. Thus, the suppression contribution of each halo population can be fully accounted for by the combination of its mass fraction and per-mass contribution to suppression.

These findings are consistent with those of \cite{Lucie_Smith_2025}, which investigates the physical processes responsible for baryonic feedback across halo mass and redshift, finding that feedback is most efficient for group-scale halos. Group-scale halos have both efficient AGN-driven feedback and relatively shallow potential wells, and contain fewer massive subhalos residing in the halo outskirts that resist AGN feedback with their own baryonic feedback expelling matter inwards towards the halo center. For a discussion on the redshift dependence of group-scale dominance, see Appendix~\ref{redshift}.

To investigate the radial dependence of baryonic feedback effects, we compare the suppression curves across the three panels of Fig.~\ref{fig: plots}. 
For group-scale halos, replacing matter within the radial range $[0.5, 2]R_{\rm 200m}$ produces little additional impact on the matter power spectrum. This behavior likely reflects the balance between feedback energetics and the depth of the gravitational potential wells in these systems, which restrict the redistribution of matter beyond the halo boundary.

Halos with $\log_{10}(M_{\rm 200m}/h^{-1}M_\odot) > 14$ show a small increase ($\sim1\%$) in the suppression of the matter power spectrum when replaced out to larger scales. This result likely indicates that AGN feedback in these high-mass systems is able to overcome the deeper potential well and eject matter out to larger distances, such that including these extant regions still has a suppressive effect on the matter power spectrum. These halos also show a consistent relative boost in the power spectrum at $k\sim2-4\,h\,{\rm Mpc}^{-1}$, likely a result of baryonic condensation and dark matter contraction in the inner regions of these halos, which increase the central densities and thus enhance power on intermediate scales corresponding to the one-halo term. 

In contrast, halos with $\log_{10}(M_{\rm 200m}/h^{-1}M_\odot) < 13$ exhibit a decrease in the contribution to suppression (a few percent) when replaced out to larger radii. In these halos, stellar feedback processes such as supernovae and stellar winds can dominate AGN feedback, so this result may be due to these processes only ejecting gas to modest distances. It is also possible that replacing out to larger radii includes gas that has fallen back into the halo, or even matter from other nearby halos in the case that the halo exists within a group or cluster. The exact amount of the change is dependent upon the exact manner or radius out to which we increase the mass in gravity-only simulation particles. In any case, it seems that the suppressive effects are concentrated in the central region of these halos. 

It is worth noting again that our exclusion of halos outlined in section \ref{Methodology} likely results in underestimates of suppression for large replacement radii. This is especially true for galaxy-scale halos that have significant populations in groups or clusters, meaning that our strict exclusion will eliminate a non-negligible amount of halos that only partially overlap in these cases. This will result in an underestimate that may dominate any increase in suppression from extending to larger scales. 

As a quantitative summary, we calculate the maximum suppression percentage for each combination of mass bin and replacement radius—corresponding to the minimum of the $P/P_{\rm GO}$ ratio shown in Fig.~\ref{fig: plots}. These suppression values, including those for intermediate half-dex mass bins, are reported in Table~\ref{tab:table 1}.

Interestingly, we find that the total suppression of matter power compared to the gravity-only case can be simply obtained as a sum of the suppressions from two individual mass bins. In Fig.~\ref{fig:adding bins}, we demonstrate that the total suppression of the matter power spectrum can be accurately reconstructed by summing the contributions from individual halo mass bins, both for adjacent and non-adjacent mass bins. The solid red and green curves show the suppression measured for halos within two different half-dex mass bins, $M_{\rm 200m}/h^{-1}M_\odot) \in [13.0, 13.5]$ and $[13.5, 14.0]$ respectively for the adjacent case and $M_{\rm 200m}/h^{-1}M_\odot) \in [12.5, 13.0]$ and $[13.5,14]$ respectively for the non-adjacent case.
The orange curves shows the resulting suppression from a simultaneous replacement. The dashed blue curves shows the cumulative suppression obtained by simply adding the suppression effects from the two mass bins. 

The close agreement between the two confirms that the overall suppression is approximately additive across halo mass bins. This behavior indicates that the baryonic redistribution in halos of different masses acts largely independently, with minimal cross-coupling between mass scales. Consequently, the total baryonic impact on the matter power spectrum can be understood as the superposition of the effects contributed by distinct halo populations. We show in Appendix~\ref{xcorr} that the lack of cross-correlations between mass bins is not guaranteed mathematically.
The near additivity of different adjacent and non-adjacent mass bins is largely expected from halo-model based arguments, where mass redistribution due to baryonic effects occur halo-by-halo as a superposition of the radial profiles of different matter components (e.g. gas, stars, collisionless matter) \cite{Schneider_2019}. In this framework, the 2-halo term --- which
accounts for the fact that halos are predominately located in high-density environments --- depends only on the dark matter component of halos. This means that baryonic effects within a halo are expected to have a negligible impact on the distribution of matter in and around other halos. \cite{Sunseri_2023} finds that most matter ejected by baryonic feedback resides in filaments, structure not directly included the traditional halo model, thereby increasing the mass of these structures and initiating further baryonic feedback processes that may potentially impose cross-correlations. However, filaments are shown to have a small impact on matter redistribution that is greatly subdominant to halo contributions, and are thus not expected to result in meaningful cross-correlations between halos. 

The expectation for cross-correlations in the case of parent halos and their contained subhalos is less clear. We find no significant cross-correlations even between cluster-scale and galaxy-scale halos, which includes galaxy-scale halos residing within parent clusters. This is expected due to the nature of our replacement scheme: in cases where a replaced halo is a subhalo of a more massive system, the impact of feedback from the host halo is baked-in to the replaced subhalo. When both bins are replaced simultaneously, the subhalo will be ignored for replacement, but will be implicitly replaced by nature of being included with the particles of the host halo. As a result, our analysis does not serve as a direct probe of the relationship between the baryonic feedback of parent halos and their hosted subhalos.

\cite{Lucie_Smith_2025} finds that feedback effects from massive subhalos counteracts mass loss in clusters by blowing gas into the Lagrangian region destined to become the cluster during its evolution, helping it to maintain and grow its mass over time. However, they also find that the strength of baryonic feedback within a halo is independent of redshift and its past or future mass (for $z<2$), depending only on its current mass, meaning that for any given snapshot this relationship will not impose cross-correlations between a parent halo and its contained subhalos. 

We note that although our choice to neglect unmatched halos systematically ignores subhalos that were significantly disrupted by baryonic feedback from their parent halos (see section \ref{Data}), these cases are very rare and are unlikely to have any meaningful impact on the contribution to power spectrum suppression.
We also note that our hard exclusion scheme will not significantly impact the detection of cross-correlations, as all particles within a halo, including those of its subhalos, are replaced, with only cases of partial overlap resulting in any neglected particles. Additionally, our findings of linear additivity between mass bins is consistent across all replacement radii, indicating that our exclusion and replacement scheme does not meaningfully impact this result. 

While we focus on the $z=0$ snapshot for our primary analysis, we also replicate Figs. \ref{fig: plots} and \ref{fig:adding bins} for the $z=0.5$ snapshot in Appendix~\ref{redshift}. We find that although suppression is slightly reduced for group- and cluster-scale halos while largely unchanged for galaxy-scale halos, the qualitative results are identical to our $z=0$ analysis. Group-scale halos still dominate suppression and the joint contribution of multiple mass bins is still a nearly linear addition of the individual bins (i.e. negligible cross-correlations between mass bins).   

\subsection{Impact on the group-galaxy weak lensing signal}

Given that our results indicate a non-monotonic dependence of baryonic effects on halo mass in the total matter power spectrum, we next ask whether this behavior can serve as an observable signature of baryonic feedback. To explore this, we turn to the cross-correlation of matter with halos of different masses as probed by the galaxy–galaxy (or group–galaxy) lensing signal. Baryonic feedback processes are expected to modify this weak lensing signal, $\Delta\Sigma(R)$, through the redistribution of matter around halos.

The excess surface density is defined as
\begin{align}
\Delta\Sigma = \bar\Sigma(<R)-\langle \Sigma(R)\rangle\,,
\end{align}
where $\bar\Sigma(<R)$ is the mean projected surface density within a projected radius $R$ from the halo center, and $\langle \Sigma(R)\rangle$ is the azimuthally averaged surface density at that same distance. The corresponding observable in weak lensing measurements is the mean tangential shear of background galaxies, $\gamma_{\rm t}$, which relates to $\Delta\Sigma$ via
\begin{align}
\gamma_{\rm t} = \frac{\Delta\Sigma}{\Sigma_{\rm crit}}\,.
\end{align}
Here, $\Sigma_{\rm crit}$ is the critical surface density, which depends on the geometry of the lensing configuration and thus on the redshifts of both lenses and sources. The critical surface density is given by
\begin{align}
\Sigma_{\rm crit} = \frac{c^2}{4\pi G} \frac{D_s}{D_l D_{ls}} \frac{1}{(1+z_l)^2}\,,
\end{align}
where $D_s$, $D_l$, and $D_{ls}$ are the angular diameter distances to the source, to the lens, and from the lens to the source, respectively. On the angular scales relevant to our analysis, the uncertainty in $\gamma_{\rm t}$ is dominated by the intrinsic ellipticity of galaxies (the so-called shape noise). 

We project the three-dimensional matter density along the line of sight toward the halos and compute $\Delta\Sigma(R)$ for the same halo mass bins used in our matter power spectrum analysis. This yields predictions for the weak lensing signal in both the full-physics and gravity-only simulations. We then quantify the impact of baryonic feedback by comparing the stacked $\Delta\Sigma(R)$ profiles of matched halos between the two simulations. Errors on these measurements are estimated assuming the depth and source density of a DES-like survey. The sample variance in each mass bin is computed via jackknife resampling, while the shape noise is estimated using
\begin{align}
\sigma_{\Delta\Sigma} = \Sigma_{\rm crit} \frac{\sigma_\varepsilon}{\sqrt{N_{\rm lenses}\,n\,\Omega}},
\end{align}
where $\sigma_\varepsilon = 0.25$ is the root-mean-square intrinsic ellipticity per galaxy \citep{Amon_2022}, $N_{\rm lenses}$ is the number of lensing halos in a given mass bin, $n = 4.2\,\mathrm{galaxies\,arcmin^{-2}}$ is the surface density of background source galaxies, and $\Omega$ denotes the solid angle of the annulus within which the tangential shears are averaged. The factor $N_{\rm lenses} n \Omega$ therefore represents the total number of lens–source pairs contributing to the measurement in each radial bin.

To emulate realistic survey conditions, we assume source galaxies distributed over $z_s \in [0.5, 1.0]$ and lens halos over $z_l \in [0.2, 0.4]$, consistent with analyses using \textsc{MagLim} lenses and DES source galaxies \citep{Amon_2022, Prat_2022}. The number of lenses in each mass bin is derived from their number density in the IllustrisTNG simulation, scaled to match the comoving volume probed by DES in this redshift range. This implicitly assumes that all central galaxies in halos within this mass range are observable at such low redshifts, a reasonable approximation given the survey depth of DES.

\begin{figure}[htbp]
    \makebox[\textwidth][c]{\includegraphics[width=\linewidth]{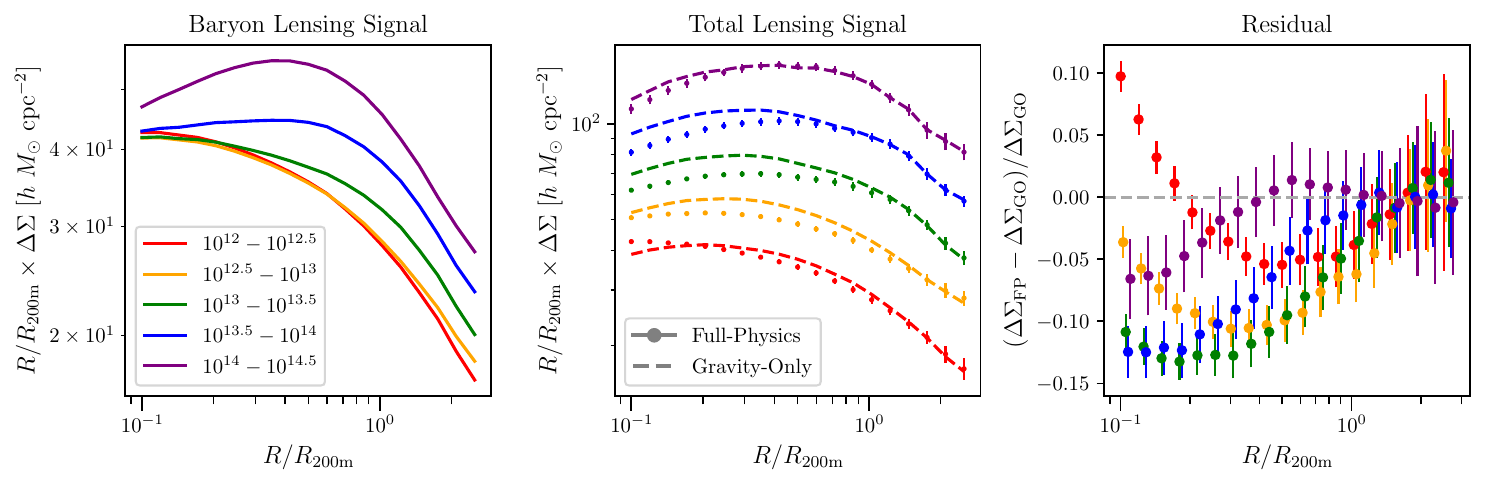}}
    \caption{\textbf{Left:} The mean weak lensing signal profile of baryons for each mass bin, i.e. the sum of the gas, star, and black hole lensing signals. \textbf{Center:} The total lensing profiles for each halo mass bin from the full-physics case with estimated shape noise and sampling error mimicking a DES-like survey, plotted alongside the lensing profiles from the gravity-only simulation. 
    Note that these first two plots are multiplied by $R/R_{\rm{200m}}$ for visual clarity. 
    \textbf{Right:} The residual profile, i.e. the difference between the gravity-only and full-physics weak lensing signals divided by the gravity-only signal for each mass bin. }
    \label{fig:lensing prof}
\end{figure}

The resulting $\Delta\Sigma(R)$ profiles are shown in Fig.~\ref{fig:lensing prof}. The left-hand panel displays the net lensing signal of baryons, combining the contributions from gas, stars, and black holes, for each halo mass bin. The distinct shapes of these profiles highlight the deviations of the baryonic components from the underlying matter distribution in the collisionless simulation.

In the middle panel, we compare the total weak lensing signal in the full-physics simulation to that in the gravity-only run. The error bars represent the combined contribution from sample variance and shape noise. This comparison isolates the impact of baryonic feedback processes on the matter distribution. Within the inner regions of halos, the baryonic profiles follow approximately $\Delta\Sigma(R) \propto R^{-1}$, consistent with near isothermal profiles for baryons. At larger radii, the profiles steepen towards $\Delta\Sigma(R) \propto R^{-2}$, reflecting the transition to the outskirts where feedback-driven redistribution of matter becomes relatively insignificant.

The left-hand panel shows the relative difference between the weak lensing profiles in the full-physics and gravity-only simulations. For the lowest-mass halos, baryonic condensation in the central regions enhances the lensing signal at scales below $\sim 0.2 R_{\rm vir}$. Beyond this radius, however, the signal becomes suppressed relative to the collisionless case due to feedback-driven mass redistribution. As halo mass increases, this central enhancement diminishes, and the maximum suppression is observed for group-scale halos with $\log M_{\rm 200m} \in [13, 14]$. At even higher masses, the suppression weakens again, consistent with the deeper potential wells that limit baryonic outflows. Beyond the virial radius, the full-physics and collisionless profiles converge, indicating that baryonic effects are confined largely within a few halo radii.

These results demonstrate that the suppression of the matter power spectrum seen in the full-physics simulation has a clear counterpart in the weak lensing signal. In particular, the scale-dependent deviations in $\Delta\Sigma(R)$ encode the same redistribution of matter responsible for the power spectrum suppression. This suggests that baryonic feedback can be effectively probed through weak lensing measurements, especially when analyzed as a function of halo mass. While an exact division of halos into narrow mass bins is challenging in observational data, practical alternatives exist. For instance, one can stack halos in bins of group luminosity, exploiting the strong correlation between total luminosity and halo mass. Abundance matching with cosmological simulations can then be used to calibrate these luminosity bins to their corresponding halo masses, allowing for an observational test of the mass-dependent baryonic effects predicted here. Another option is to use the 1-halo term of the measured weak lensing signal to get the estimate of the mass of the halo directly and study the redistribution of matter within halo of given mass. Both options have pros and cons. While the former approach relies on the intermediate group properties and abundance matching, it will be able to measure both shape and amplitude change in the density profile. The latter does not involve going through abundance matching and is a more direct observational probe, although that will only be able to probe the change in the shape of the density profile. 

We note that \cite{hadzhiyska2025} found that IllustrisTNG systematically underestimates baryonic feedback effects compared to what is observed in kSZ selected clusters. This points towards potentially even stronger deviation from the gravity-only weak lensing signal at small scales, and some of the conclusions may depend upon the exact implementation. However, the deviation of the weak lensing profiles from expectations from collisionless simulations can be a useful probe of the baryonic effects and distinguish it from changes in cosmology.

\section{Summary and Discussion}
\label{discussion}

In this work, we have developed and applied a hybrid simulation framework to quantify the impact of baryonic feedback on the matter distribution using the IllustrisTNG simulations. By selectively replacing particles in the collisionless TNG300-1-Dark simulation with their full-physics counterparts as a function of halo mass and radial extent, we isolated how baryonic processes modify structure formation across different mass scales. Our key results can be summarized as follows:

\begin{itemize}
\item We constructed controlled realizations of the matter field by replacing particles within spherical regions of radius $\alpha R_{\rm 200m}$ around halos in different mass bins and matched between the full-physics and gravity-only simulations. Variations in the halo mass bins and $\alpha$ allowed us to study the mass and radial dependence of the baryonic effects on the matter power spectrum.
\item Nearly 90 percent of the total suppression of the matter power spectrum in the full-physics simulation, which reaches up to 20\% at $k \sim 10\,h\,\mathrm{Mpc}^{-1}$, can be explained due to feedback from halos with masses $\log M_{\rm 200m} > 12$.
\item Group-scale halos ($\log M_{\rm 200m} \in [13,14]$) contribute the dominant $\sim10$ percent suppression, representing 60 percent of the total effect from all halos above $\log M_{\rm 200m}=12$. Lower-mass halos ($\log M_{\rm 200m} \in [12,13]$) and cluster-scale halos ($\log M_{\rm 200m} > 14$) each contribute only a few percent to the total suppression. The effect is non-monotonic with halo mass, it peaks at group scales, consistent with efficient AGN feedback redistributing matter beyond the virial radius. 
\item We find that the dominant contribution to matter power suppression by group-scale halos can not be solely attributed to their large total mass fraction within the simulation. Group scale halos also host more efficient feedback, contributing nearly double to suppression on a ``per-mass" basis than both galaxy- and cluster-scales. 
\item We demonstrated that the suppression from different halo mass bins adds linearly, demonstrating that baryonic feedback operates as a near-independent contribution from each mass scale.
\item Effects of feedback on the matter power spectrum can be simply explained by redistribution of matter within $R_{\rm 200m}$ and redistributing some fraction of this mass expelled by the feedback effects into the surrounding volume at a distance about $1.5$ times larger.
\item The differential surface density profiles, $\Delta\Sigma(R)$, from the full-physics simulation show enhanced central signals at low masses and suppressed signals at larger radii, mirroring the redistribution seen in the power spectrum. The maximum suppression in $\Delta\Sigma$ occurs for group-scale halos, again highlighting its importance in baryonic feedback.
\end{itemize}

Our results suggest that mass-dependent weak lensing measurements, particularly when stacking by group luminosity or richness, can directly probe baryonic feedback signatures. The distinct non-monotonic dependence of suppression on halo mass provides a potential observational discriminator of feedback models. Although seemingly straightforward, some systematics are worth pointing out. Group finding algorithms often have trouble identifying isolated halos, and could either fragment groups, or combine them due to line-of-sight projection effects (see e.g., \cite{Campbell2015}). There are also potential issues related to the identification of central galaxies, as well as the mis-centering between central galaxies and the true potential centers of halos. We defer explorations of these systematics to future work.  

Our conclusions are largely consistent with those of \cite{vanloon_2023}, in which the impact of AGN feedback was quantified across halo mass scales in the cosmo-OWLS and BAHAMAS simulation projects. Their study compared the power spectrum of particles within the spherical overdensity regions (both $R_{\rm{200m}}$ and $R_{\rm{500c}}$) of full-physics halos to that of the entire full-physics simulation, the gravity-only simulation, and matched gravity-only halo particles. 
They find that the contribution of halo mass bin to the matter power spectrum suppression is related to the fraction of mass redistributed beyond the virial radius. Similar to our findings, they identify group-scale halos as major contributors and both low- and high-mass systems as contributing subdominantly. 

Quantitative suppression levels derived from simulations, including total suppression and the scale at which suppression peaks, depend on the specific subgrid feedback model implemented within the simulations. IllustrisTNG implements the moving-mesh \textsc{AREPO} code \cite{Springel_2010, Weinberger_2020} including prescriptions for star formation, stellar evolution, chemical enrichment, primordial and metal-line cooling, stellar feedback with galactic outflows, and black hole formation, growth and multi-mode feedback \cite{Pillepich_2017}. IllustrisTNG's updated thermal and kinetic AGN feedback prescription allows energy injection at low accretion rates, and yields excellent agreement with both simulated and observed stellar mass fractions without overheating gas \cite{Weinberger_2016}. 
On the other hand, the BAHAMAS simulation project (and its predecessor cosmo-OWLS) uses the hierarchical tree algorithm \textsc{GADGET}, and is designed specifically to probe the impact of baryonic processes on large-scale correlations, offering a large box-size of $400\,h^{-1}\,\rm{Mpc}$ (compared to IllustrisTNG's $205\,h^{-1}\,\rm{Mpc}$). Notably, BAHAMAS attempts to calibrate feedback processes on galaxy properties as well as the integrated gas fractions of galaxy groups and clusters \cite{Chisari_2019}. 

Both treatments of baryonic feedback produce a relative suppression of the matter power spectrum at small scales, but disagree on the exact effect. Suppression in BAHAMAS begins at $k\sim0.5\,h\,\rm{Mpc}^{-1}$ and hits its peak of around $15\%$ at $k\sim5\,h\,\rm{Mpc}^{-1}$. In IllustrisTNG, suppression begins later at $k\sim2\,h\,\rm{Mpc}^{-1}$, bottoming out with about 20\% suppression at $k\sim10\,h\,\rm{Mpc}^{-1}$. 
OWLS has the greatest total suppression, beginning at $k\sim0.5\,h\,\rm{Mpc}^{-1}$ and reaching a maximum greater than 20\% at $k\sim10\,h\,\rm{Mpc}^{-1}$.
See \cite{Chisari_2019} for an in-depth comparison between these and other simulations. 
This disagreement indicates a significant difference in the treatment of baryonic effects between all three methodologies. However, our agreement with the analysis of \cite{vanloon_2023} using BAHAMAS and cosmo-OWLS suggests that these differences do not significantly impact conclusions drawn about the relative contribution of different mass halos to matter power suppression, or the finding that the halo mass fraction is tightly connected to this relative contribution. 

In general, our results are largely robust to the subgrid feedback model chosen by a simulation. Varying feedback strength or coupling efficiency will result in slight shifts in the mass range contributing most dominantly to suppression and the scales at which suppression begins and peaks. However, the qualitative result that group-scale halos dominate is expected to be robust because it largely reflects the underlying halo mass function weighting in combination with AGN efficiency. This is evidenced by our consistent results with analysis using cosmo-OWLS, which predicts stronger AGN feedback than IllustrisTNG. We expect this to extend to even stronger AGN feedback models, such as in the SIMBA simulation suite \cite{Dav__2019}, as well as weaker feedback, such as in the EAGLE \cite{McAlpine_2016} and Horizon-AGN \cite{Dubois_2014} simulations.

\subsection{Connections to weak lensing and SZ measurements}

We have shown that the suppression of the matter power spectrum directly translates into weak-lensing observables such as the cosmic shear power spectrum and the galaxy-group lensing signal \cite{Chisari_2018, Chisari_2019}. Our results therefore underscore the need to model the halo mass dependence of baryonic feedback in future weak lensing analyses. While current 3$\times$2-point analyses often exclude small scales due to feedback-related uncertainties \cite{Secco_2022}, properly incorporating these scales, using physically motivated models or simulation-based calibrations, could recover a significant amount of cosmological information currently discarded \cite{Huang_2021, Chen_2022, Aric__2023, Schneider_2022}.

The peak contribution to cosmic shear occurs at $z \simeq 0.5$, somewhat higher than the redshift studied here. While our results at $z=0$ capture the cumulative impact of baryonic processes, the redshift evolution of the suppression is expected to be mild for $z \lesssim 1$ and $k \lesssim 10,h,\mathrm{Mpc}^{-1}$ \cite{Schneider_2019}. We show in Appendix~\ref{redshift} that our key results are unchanged at $z=0.5$, where the suppression contribution is slightly reduced for group- and cluster-scale halos. This indicates that our results are robust to the redshift scales typically of interest for weak lensing analyses. Nonetheless, the joint dependence of feedback on halo mass, spatial scale, and cosmic time remains an open question that future simulations and tomographic weak lensing analyses could address.

Complementary probes of baryonic feedback are emerging from observations of the thermal and kinetic Sunyaev–Zel’dovich (tSZ and kSZ) effects in groups and clusters \cite{To_2024, hadzhiyska2025, Lucie_Smith_2025, bigwood_2024, hadzhiyska_2023, roper_2025}. These measurements directly trace the thermodynamic state of the gas, providing independent constraints on the redistribution of baryons within and around halos.
For instance, \cite{roper_2025} use kSZ measurements to constrain the mass-dependent halo baryon fraction, finding evidence for extended gas profiles beyond the virial radius in halos of mass $10^{12.5} \lesssim M/h^{-1}M_\odot \lesssim 10^{14}$, supporting our findings that these halos redistribute significant mass beyond the virial radius. 
The group-mass regime, which our work has identified here as most influential for the matter power suppression, is also the most accessible to current kSZ and weak lensing measurements. While most SZ observations were limited to relatively massive systems, recently stacked SZ–lensing analyses have been shown to extend sensitivity to lower masses, modulo uncertainties in the halo–observable relation \cite{Lin_2021, zhang_2023, bigwood_2024}. 
Forthcoming high-sensitivity CMB experiments, such as the Simons Observatory \cite{SO_2025}, promise to probe baryonic feedback across a broader halo mass range and provide crucial constraints on the rearrangement of matter in the Universe.

Finally, while baryonic probes such as X-ray and SZ measurements are directly sensitive to the baryonic gas, we emphasize that the total matter distribution, i.e., baryons and the dark matter, is jointly rearranged by feedback. This redistribution alters the gravitational potential and thus affects cosmological observables even where baryons themselves are subdominant. Surveys such as the Rubin LSST which will start imminently will increase the surveyed area compared to the Dark Energy Survey by a factor of 4, and the better depth will lead to an increase in the number of galaxies per square arcmin of at least 3. This will significantly reduce the statistical error on the cosmic shear measurements down to small scales, as well as the group galaxy lensing signals. Together with the exquisite angular resolution and depth provided by space based surveys such as Euclid and Roman, the quality and quantity of data will increase dramatically in the near future. To realize the scientific output of these surveys to the fullest, joint interpretation of future high-precision weak lensing, X-ray and SZ data will therefore require models that capture both the baryonic and the induced dark matter response to feedback.

\subsection{Improving Theoretical Models} \label{theory models}

The results presented in this paper demonstrate that baryonic feedback produces a non-monotonic and mass-dependent impact on both the matter distribution within halos and the total matter power spectrum. This complexity highlights a key challenge for cosmological inference: baryonic correction models and emulators should not just be tuned to reproduce the total matter power spectrum. Our findings show that baryonic feedback reshapes the internal structure and outskirts of halos in a mass-dependent manner. Accurately capturing this may require a new generation of emulators capable of jointly predicting the halo–matter connection and the corresponding large-scale matter correlation in a self-consistent framework.

Emulators such as \textsc{DarkEmulator} \cite{Nishimichi_2019} and \textsc{BACCOemu} \cite{Aric__2021} provide predictions of nonlinear clustering and halo statistics, but do not include both small scales and halo matter cross correlations in the presence of baryonic effects. Hybrid emulators which can parameterize the mass dependent redistribution of matter within halos such as HMx \cite{Mead2020} or the baryonification project \cite{Osato2022} which could be adapted to self-consistently predict the halo matter cross-correlations and the impact on the matter power spectrum are promising ways forward.

Semi-analytic phenomenological methods of painting the effects of baryons onto N-body simulations ("baryonification") are becoming ubiquitous tools for modeling structure formation at non-linear scales. These methods offer a fast and computationally cheap alternative to intensive hydrodynamical simulations, especially useful for training emulators and estimating covariance, processes which requires hundreds or thousands of large-volume simulations with varying cosmological and baryonic feedback parameters. Such methods typically use halo-model based prescriptions, where halo density profiles are broken down into their constituent dark matter, gas and stellar components, to redistribute particles in and around halos in an N-body simulation (e.g. \cite{Mead2020, Osato2022, Schneider_2019, Anbajagane_2024}). This approach has also been adapted to baryonify 2D lightcone maps from simulations, redistributing pixel values instead of particle positions (e.g. \cite{Fluri_2019, Anbajagane_2024}). These methods baryonify simulation results halo-by-halo according to both a 1-halo term containing the multi-component density profile and a 2-halo term, which is a component only of the dark-matter-only profile \cite{Schneider_2019}. Given the halo-based prescription of baryonic effects in these methods, our results have direct implications for both their modeling and parameter tuning. 

The slope of the gas profile as a function of halo mass (governed by parameter $M_{\rm{c}}$) and the distance out to which gas is ejected ($r_{\rm{ej}}$) are important parameters in baryonification models. Our results provide insight into the distance out to which feedback is efficient in redistributing matter around halos of different masses, thus they may be useful in informing these parameters.
 
Additionally, our finding of negligible cross-correlations between baryonic feedback from halos in different mass bins in the suppression of the matter power spectrum supports the halo-by-halo treatment of the baryonification methods, with cross correlations only potentially stemming from the 2-halo term, which contains no dependence on baryonic physics. 
We also show that the contribution of filaments is greatly sub-dominant to that of halos, however, as greater precision is demanded by ever-improving datasets, the inclusion of filaments into future baryonification schemes may become worthwhile. 

Measurements of baryonic effects from visible tracers (e.g. FRBs, SZ) will be essential in calibrating the aforementioned modeling for cosmological analysis. \cite{vanloon_2023} present a ``resummation" model (implemented and expanded upon in \cite{vandaalen_2025}) for the suppression of matter clustering due to baryonic effects. This model uses the baryon fractions within halos of a given mass to estimate suppression of the matter power spectrum. In parallel to our results, they show with hydrodynamical simulations that the expected mass deficit within spherical overdensity regions of radii $R_{\rm{200m}}$ and $R_{\rm{500c}}$ due to the expulsion by baryonic effects can be used to predict the power spectrum suppression. Such models, while computationally expensive, offer robust non-parametric treatment of baryonic effects using visual tracers that will become even more precise with upcoming next-generation surveys and can be used in combination with lensing to jointly constrain cosmological parameters as well as baryonic feedback effects. 

\section*{Acknowledgements}
We would like to thank Arka Banerjee, Susmita Adhikari, Sunao Sugiyama, Shivam Pandey, Dhayaa Anbajagane, Shubham Sati, and Shreya Mukherjee for the helpful feedback and discussions, and Dylan Nelson for his help in providing access to the Illustris TNG simulation and the python notebook server close to the data. We would also like to thank the referee for their careful reading of the manuscript and their comments which helped improve the paper and its presentation. KM would like to thank IUCAA for its hospitality during his visit to India where a large fraction of this research work was conducted.  We would also like to thank Marcel van Daalen, Yan-Chuan Cai, and Luisa Lucie-Smith for feedback on an earlier version of the manuscript.

The IllustrisTNG simulations were undertaken with compute time awarded by the Gauss Centre for Supercomputing (GCS) under GCS Large-Scale Projects GCS-ILLU and GCS-DWAR on the GCS share of the supercomputer Hazel Hen at the High Performance Computing Center Stuttgart (HLRS), as well as on the machines of the Max Planck Computing and Data Facility (MPCDF) in Garching, Germany. KM and BJ are partially supported by the US Department of Energy grant DE-SC0007901 and by NASA funds for the Roman Space Telescope PIT. 

\appendix
\section{Mass Conservation} \label{masscons}

\begin{figure}
    \centering
    \includegraphics[width=\linewidth]{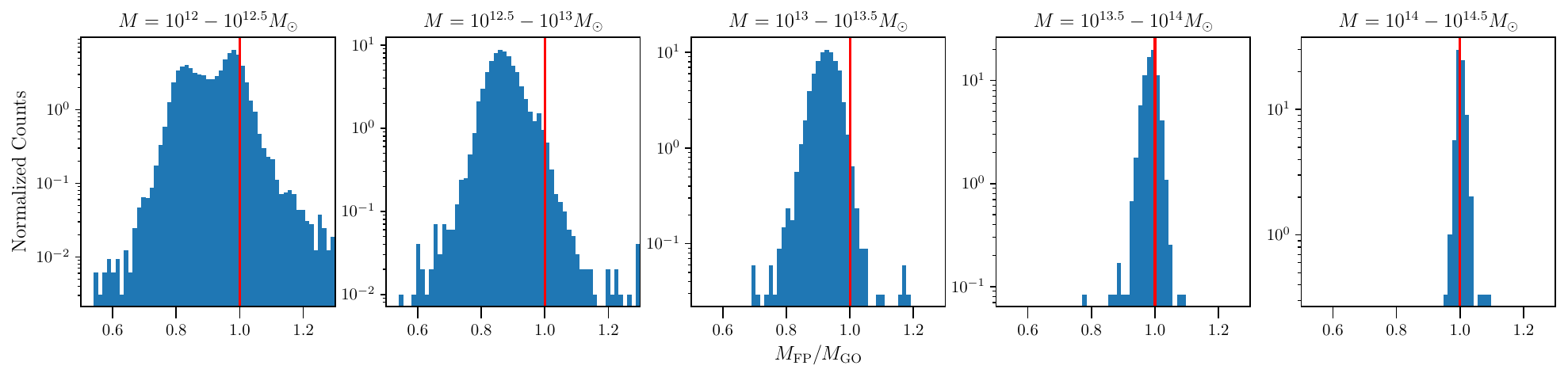}
    \caption{The histogram of mass ratios between corresponding halos in the full-physics and gravity-only simulations. Full-physics halos are seen to generally have a slightly lower $M_{\rm{200m}}$ relative to their gravity-only simulation counterparts because baryonic feedback processes eject gas ejected out beyond $R_{\rm{200m}}$.}
    \label{fig:mass ratio}
\end{figure}

We replace matter in the gravity-only simulation with that from the full-physics simulation. When halos are matched between the two, we find that the total replaced mass does not exactly equal the original mass in the gravity-only case. This difference is expected as baryonic feedback processes such as AGN activity and supernovae eject gas to large radii, while gas cooling and galaxy formation pull matter inward. Consequently, the mass enclosed within a fixed spherical radius around a halo center differs between the two simulations. Typically, full-physics halos contain less mass than their gravity-only counterparts, except at the highest halo masses where deeper potential wells retain more baryons. This trend is evident in Fig.~\ref{fig:mass ratio}, which shows the ratio of $M_{\rm 200m}$ between the matched halos.

\begin{figure}
    \centering
    \includegraphics[width=\linewidth]{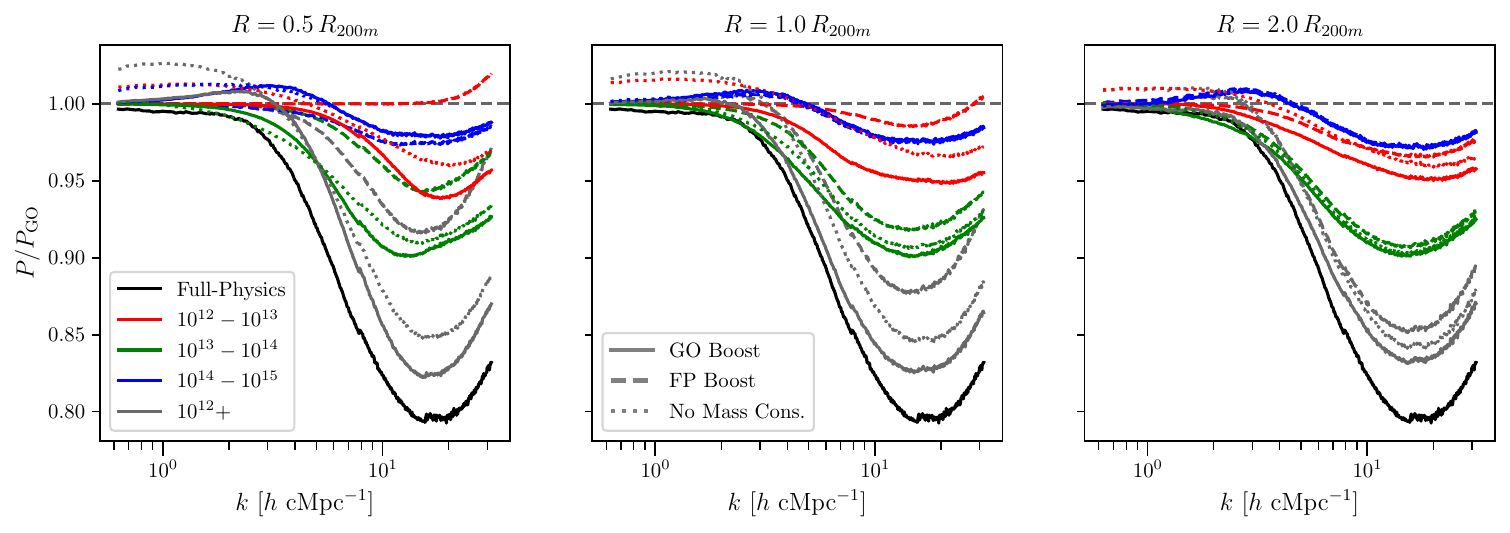}
    \caption{$P/P_{\rm{GO}}$ for each non-overlapping one-dex mass bin for three different mass conservation methods. Solid curves correspond to halo-by-halo redistribution of the mass deficit to surrounding gravity-only simulation particles. Dashed curves correspond to halo-by-halo boosting of the replaced full-physics particles. Dotted curves correspond to no mass conservation at all.}
    \label{fig:mc comp}
\end{figure}

When halo particles are replaced within a fixed multiple of $R_{\rm 200m}$, the result is generally a modest reduction in total halo mass. Summed over the full volume, these losses correspond to a global mass deficit of order $\sim1\%$. While this reduction is physically meaningful, if unaccounted for, it can introduce a spurious boost in large-scale power, as seen in the dashed curves of Fig.~\ref{fig:mc comp}. The effect is strongest for the lowest halo mass bins ($M_{200m}\in[10^{12},10^{13}],M_\odot$) and for small replacement radii ($\alpha\in[0.5,1.0]$). At larger radii or when replacing more massive halos, the boost becomes negligible because the replaced regions encompass more of the baryon-rich outskirts expelled by feedback, reducing the net mass deficit. Physically, the loss of mass on small scales lowers the local overdensity contrast and redistributes power into larger-scale modes, thereby producing the apparent increase in large-scale clustering amplitude.

We implement and compare two complementary methods of enforcing mass conservation, both of which successfully eliminate the spurious large-scale power boosts, but differ in their impact on small-scale structure. In the first approach, we locally rescale the masses of replaced particles within each halo by a factor of $M_{\rm GO}(<R) / M_{\rm FP}(<R)$, where $M_{\rm GO}(<R)$ and $M_{\rm FP}(<R)$ are the enclosed masses within the replacement radius in the gravity-only and full-physics simulations, respectively. This results in a mere change in the normalized density profile of a halo, and ensures that each halo retains its original mass after replacement. Such replacement removes the large-scale boost (see dashed lines in Fig.~\ref{fig:mc comp}). However, the rescaling also counteracts the very baryonic effects we aim to capture, i.e., the expulsion of matter from halo interiors to larger radii. As a result, the small-scale suppression in the power spectrum is artificially underestimated. Although some residual suppression remains due to the altered spatial distribution of particles, this method does not fully represent the physical redistribution of baryons.

Our second, more physically motivated approach conserves both halo and global mass by redistributing the missing mass to the gravity-only particles surrounding each halo, rather than rescaling particles inside the replacement region. Specifically, the mass deficit is distributed uniformly among gravity-only particles within a spherical shell extending from $R_{\rm replace}$ to $2R_{\rm replace}$. This mimics the baryonic feedback process, where ejected material populates the outskirts of halos. This approach also removes the large-scale boost, but results in a somewhat stronger small-scale suppression than the local rescaling method. This enhancement is physically plausible, as baryonic feedback indeed transfers mass from the halo interior to outer regions, lowering central densities. By distributing mass outward rather than scaling it up, this scheme preserves the qualitative effect of feedback while enabling a controlled study of how different radial ranges contribute to the total suppression of the matter power spectrum (see solid lines in Fig.~\ref{fig:mc comp}). We have experimented with various factors out to which we redistribute the mass, and find our conclusions robust to our fiducial choice. 

\section{Redshift Dependence} 
\label{redshift}

\begin{figure}
    \centering
    \includegraphics[width=\linewidth]{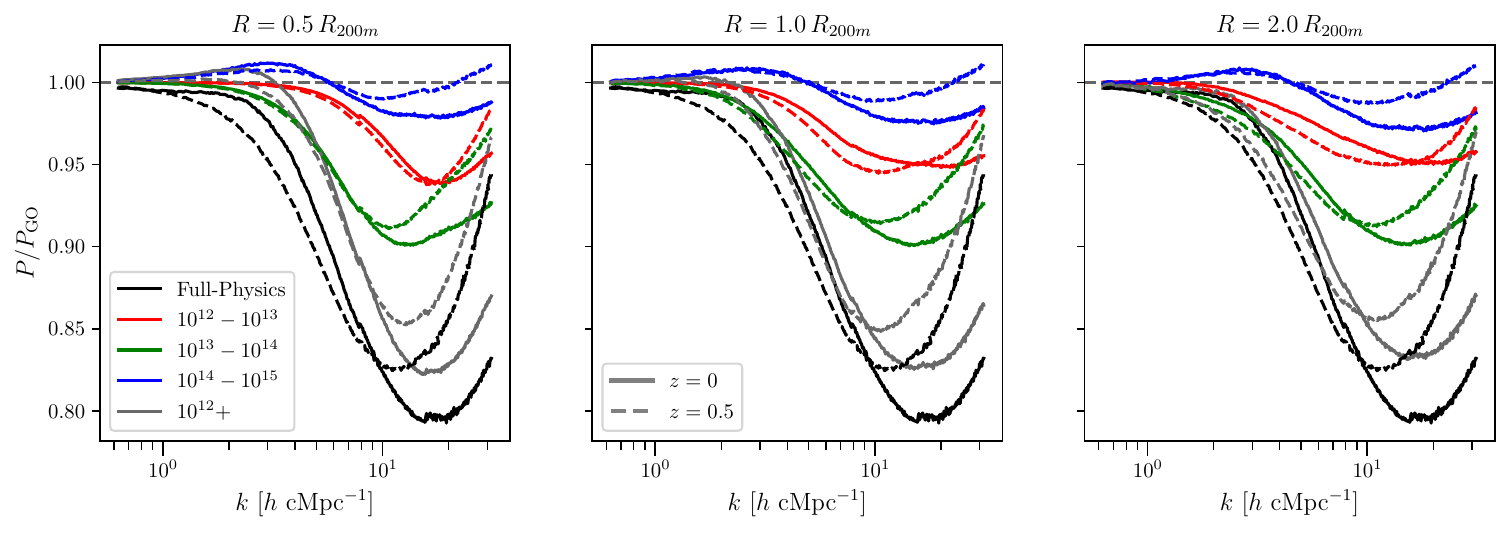}
    \caption{Replication of Fig.~\ref{fig: plots} including both the $z=0$ and $z=0.5$ results. The solid curves correspond to the $z=0$ results presented in Section~\ref{results}, while the dashed curves correspond to the $z=0.5$ results.}
    \label{fig: plots z comp}
\end{figure}

\begin{figure}
    \centering
    \includegraphics[width=0.65\linewidth]{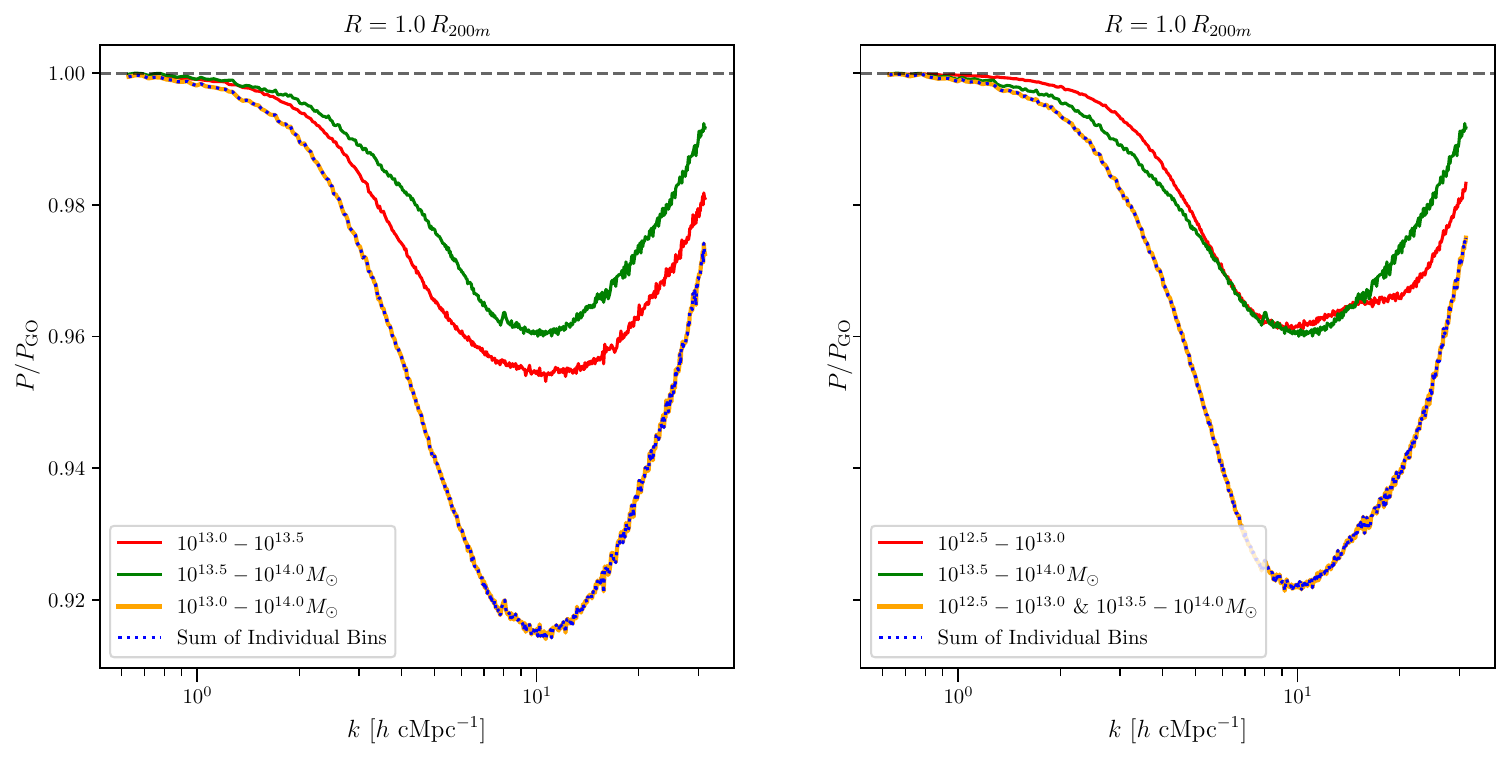}
    \caption{Replication of Fig.~\ref{fig:adding bins} for the $z=0.5$ snapshot. Once again the suppression contributions of individual bins add nearly exactly linearly to get the contribution from joint replacement, indicating negligible cross correlations for both adjacent and non-adjacent mass bins.}
    \label{fig:adding bins z=0.5}
\end{figure}

In order to ensure that our results are robust to the redshifts most relevant to weak lensing analyses ($z\simeq0.5$), we reproduce Figs. \ref{fig: plots} and \ref{fig:adding bins} for the $z=0.5$ snapshot. We plot the power spectrum ratios across replacement radii for each mass bin in both redshift snapshots in Fig.~\ref{fig: plots z comp}, and the addition of power spectrum ratios compared to joint replacement in Fig.~\ref{fig:adding bins z=0.5}. 

We find that the shape and depth of the curves differ slightly between mass bins, with all mass bins reaching their peak suppression at slightly lower $k$ than for $z=0$, and group- and cluster- scales having slightly reduced suppression contributions corresponding with an overall reduced suppression for the entire full-physics simulation. Despite these differences, the relative contribution of each mass bin remains largely unchanged, with galaxy-scales contributing a slightly larger fraction than for $z=0$, but with group-scales still dominating the effect. We also find that the lack of cross-correlations between mass bins persists for $z=0.5$, with near exact linear additivity of contributions to matter power suppression from different mass bins. 

\cite{Lucie_Smith_2025} investigates the underlying physical processes behind baryonic feedback across halo mass and redshift. They find that feedback efficiency depends only on the instantaneous mass of the halo at a given redshift, with no dependence on past or future mass for $z<2$. In parallel with our results, they find that group-scale halos have the most efficient feedback across these redshifts. 
Additionally, halos with mass $M_{\rm{200m}}>10^{14}M_\odot$ were found to have less effective feedback at late times ($z<2$) despite significant mass loss due to feedback at early times ($z=2-4$). This is due to both re-accretion of ejected baryons due to their strong gravity and matter ejected into the halo center by massive subhalos residing within, both of which restore these cluster-scale halos to their pre-feedback mass. For the scales most relevant to weak lensing, however, the redistribution of matter due to baryonic feedback for a given halo mass is consistent across all redshifts.

\section{Cross-Correlations} \label{xcorr}

We show that the linear additivity of suppression contributions from each mass bin is not guaranteed mathematically. Let us consider the two-point correlation function estimator
\begin{align}
\xi(r) = \frac{DD(r)}{RR(r)} - 1\,,
\end{align}
where
$DD(r)$ is the number of particle pairs separated by $r$, and $RR(r)$ is the number of random pairs.

We can split the particles into three disjoint sets \{$A$, $B$, $C$\}, where $A$ denote particles in halos belonging to mass bin 1, $B$ denotes particles in halos belonging to mass bin 2, and $C$ denote particles in all other halos. Every particle pair thus belongs to one of six categories, $AA,\ BB,\ CC,\ AB,\ AC,\ BC$. Let us now define pair counts, $DD^{X}_{ij}$, where $X$ indicates which halo populations are replaced $(i,j)$ indicate the six categories listed above. Thus, $X=0$ would correspond to the gravity only simulation, $X=A$ will mean particles in halo mass bin 1 are replace, $X=B$ would mean particles in halo mass bin 2 are replaced, and $X=AB$ would mean particles in both bins are replaced. Given that we are enforcing mass conservation would imply that we do not need to worry about $RR$ term and can solely focus on the total paircounts.

The total pair counts in the gravity only simulation will be given by
\begin{align}
DD^{0} =
DD^{0}_{AA}
+DD^{0}_{BB}
+DD^{0}_{CC}
+DD^{0}_{AB}
+DD^{0}_{AC}
+DD^{0}_{BC}\,.
\end{align}
Once we replace the particles in halo mass bin 1, paircounts containing particles from set $A$ change. Thus we have,
\begin{align}
DD^{A} =
DD^{A}_{AA}
+DD^{0}_{BB}
+DD^{0}_{CC}
+DD^{A}_{AB}
+DD^{A}_{AC}
+DD^{0}_{BC}\,.
\end{align}
Now let us define
\begin{align}
\Delta DD_A = DD^{A}-DD^{0}\,.
\end{align}
Then, we obtain
\begin{align}
\Delta DD_A =
(DD^{A}_{AA}-DD^{0}_{AA})
+
(DD^{A}_{AB}-DD^{0}_{AB})
+
(DD^{A}_{AC}-DD^{0}_{AC})\,,
\end{align}
resulting in three terms, as the other terms do not involve any particles in set $A$. Similarly replacing particles in halos in mass bin 2, would yield,
\begin{align}
\Delta DD_B =
(DD^{B}_{BB}-DD^{0}_{BB})
+
(DD^{B}_{AB}-DD^{0}_{AB})
+
(DD^{B}_{BC}-DD^{0}_{BC})\,.
\end{align}
while replacing in both bins would result in
\begin{align}
\Delta DD_{AB} =&
(DD^{AB}_{AA}-DD^{0}_{AA})
+
(DD^{AB}_{BB}-DD^{0}_{BB})
+
(DD^{AB}_{AB}-DD^{0}_{AB}) \nonumber \\
&+
(DD^{AB}_{AC}-DD^{0}_{AC})
+
(DD^{AB}_{BC}-DD^{0}_{BC})\,.
\end{align}

Adding the effects from replacing each bin individually, we obtain
\begin{align}
\Delta DD_A+\Delta DD_B &=
(DD^{A}_{AA}-DD^{0}_{AA}) \\
&+
(DD^{B}_{BB}-DD^{0}_{BB}) \\
&+
(DD^{A}_{AB}-DD^{0}_{AB}) \\
&+
(DD^{B}_{AB}-DD^{0}_{AB}) \\
&+
(DD^{A}_{AC}-DD^{0}_{AC}) \\
&+
(DD^{B}_{BC}-DD^{0}_{BC})\,.
\end{align}
Given that we observe additivity in the suppression effect, $\Delta DD_A+\Delta DD_B=\Delta DD_{AB}$. 

The term $DD^{AB}_{AX}$ and $DD^{A}_{AX}$ where $X\in{A, C}$ should be the same, as these paircounts should not be affected by particles being replaced in mass bin 2. These correspond to $AA$ and $AC$ pairs. By a similar argument pairs correspond to $BB$ and $BC$ should not be affected by replacement of particles in set $A$. So the only terms that remain correspond to the cross-pairs $AB$. Demanding additivity requires
\begin{align}
DD^{AB}_{AB}
=
DD^{A}_{AB}
+
DD^{B}_{AB}
-
DD^{0}_{AB}\,.
\end{align}

Additivity of the suppression is therefore not guaranteed mathematically. But if baryonic feedback mainly redistributes matter locally, and the halo centers and halo-halo separations remain nearly unchanged, then we expect the additivity to hold. In our methodology we do maintain the halo separations and halo centers by construction, and also adopt the local redistribution. So it is perhaps not entirely unexpected.

\bibliographystyle{JHEP}
\bibliography{references}

\end{document}